\documentclass[10pt,twocolumn,times,superscriptaddress,showpacs,showkeys]{revtex4-1}

\usepackage{color}
\usepackage{graphics}
\usepackage{graphicx}
\usepackage{subfig}
\usepackage{amsmath}
\usepackage{amsfonts}
\usepackage{amssymb}

\begin{document}
\title{Fractional Generalized Langevin Equation Approach to Single-File Diffusion}

\author{C.H. Eab}
\email{Chaihok.E@chula.ac.th}
\affiliation{Department of Chemistry,
Faculty of Science, Chulalongkorn University,
Bangkok 10330, Thailand
}

\author{S.C. Lim }
\thanks{corresponing author}
\email{sclim47@gamail.com}
\affiliation{10E Braddell Hill, {\#}22-19, Braddell View 
Singapore 579724
}

\date{\today}

\begin{abstract}
Fractional generalized Langevin equation with external force is used to model single-file diffusion. 
It is found that for external force that varies with power law the solution 
for such a fractional Langevin equation gives the correct short and long time behavior for the mean square displacement of single-file diffusion 
when appropriate choice of parameters associated with fractional generalized Langevin equation are used. 
By considering some special cases of the fractional generalized Langevin equation, 
a new class of closed analytic expressions for the mean square displacement of single-file diffusion can be obtained.
The effective Fokker-Planck equation associated with single-file diffusion is briefly considered.

\end{abstract}


\maketitle

\section{Introduction}
\label{sec:introduction}
\noindent
Recent advances in nanofabrication allow the preparation of new types of nanotube materials such as carbon nanotubes, petide nanotubes, 
inorganic and organic zeolites, etc. 
The study of the molecular transport processes inside these nanotubes or channels have attracted considerable attention  
\cite{KargerRuthven09,Karger08,RoqueMalherbe07,ChengBowen07,Strook00,AidleyStandfield96,Alberts08}. 
The one dimension transport process of an assembly of non-passing particles (without mutual passage) 
in confined geometries such as narrow pores or nanotubes is known as single-file diffusion (SFD). 
In other words, particles undergoing SFD maintain the order of their arrangement at all times.

The main feature of SFD can be characterized by its short time and long time limits of its mean square displacement (MSD) denoted by $R^2$. 
For very short observation time, the particles diffuse normally and satisfy Fick's law, 
such that the short time limit of MSD of SFD is given by
\begin{align}
  R^2 & \equiv \left<\left(x(t)-x_\circ\right)^2\right> = 2D_\circ t,
    \quad t \rightarrow 0,
\label{eq:introduction_01}
\end{align}
where
$D_\circ$
is the diffusion coefficient. 
In other words, for very short diffusion time, 
the motion is just ordinary Brownian motion, 
which is a Markov process. 
As for the long time limit of the MSD for SFD, one has          
\begin{align}
  R^2 & = 2F\sqrt{t}, 
      \quad t \rightarrow \infty,
\label{eq:introduction_02}
\end{align}
where $F$ is the SFD mobility. 
Recall that diffusion that does not satisfy Fick's law is known as anomalous diffusion with MSD satisfying
$R^2 \propto t^\alpha$, 
$\alpha\neq 1$. 
When  
$\alpha > 1$
the diffusion is enhanced and it is called superdiffusion. 
In the case with
$\alpha < 1$
the diffusion is subdued and one has subdiffusion. 
Thus, the long time behavior of SFD belongs to subdiffusion, 
which is a non-Markovian process, indicating the motion is correlated.

Consider the simple case of a single molecule in a one-dimensional pore, 
such that the molecule can move in either direction with equal probability via activated jumps with step length 
$l$. 
If
$\tau$   
is the average time between successive jumps, 
and   
$\theta$
is the fractional occupancy, 
then one has
\begin{align}
  R^2 & = l^2(1-\theta)\frac{t}{\tau}, \quad
   t  \rightarrow 0,
\label{eq:introduction_03}
\end{align}
which corresponds to normal self diffusion. 
Fedders 
\cite{Fedders78}
has derived the long time limit of the MSD as 
\begin{align}
  R^2 & = l^2 \frac{(1-\theta)}{\theta}\sqrt{\frac{2}{\pi}}\sqrt{\frac{t}{\tau}}, \quad
    t  \rightarrow \infty,
\label{eq:introduction_04}
\end{align}
which shows that at longer times the motion is strongly suppressed by collisions with neighboring particles 
(see also \cite{Karger92}). 
From (\ref{eq:introduction_02}) and (\ref{eq:introduction_04}) one obtains the SFD mobility as
\begin{align}
F & = l^2 \frac{(1-\theta )}{\theta}\frac{1}{{\sqrt{2\pi}}}.
\label{eq:introduction_05}
\end{align}

About a decade ago, 
Brandani \cite{Brandani96} introduced heuristically (without derivation) the following analytic expression for the MSD in SFD:
\begin{align}
  R^2 & = l^2 \frac{(1-\theta)t/\tau}{1+\theta\sqrt{\pi/2}\sqrt{t/\tau}}.
\label{eq:introduction_06}
\end{align}
Expression 
(\ref{eq:introduction_06})
is able to give the correct limiting cases 
(\ref{eq:introduction_03})
and 
(\ref{eq:introduction_04}). 
Even until today, 
(\ref{eq:introduction_06})
is still regarded as an expression that 
"comprises both cases with satisfactory accuracy" 
[Ref. \cite{Karger08}, page 334].  

A similar expression for the MSD of SFD was obtained by 
Lin et al 
\cite{Lin05}
based on the following ansatz:
\begin{align}
  \frac{1}{R^2} & = \frac{1}{2D_\circ t} +  \frac{1}{2F\sqrt{t}} .
\label{eq:introduction_07}
\end{align}
By solving
(\ref{eq:introduction_07})
for MSD one gets
\begin{align}
  R^2 & = \frac{2D_\circ t}{1+D_\circ\sqrt{t}/F} .
\label{eq:introduction_08}
\end{align}
Just like  
(\ref{eq:introduction_06}), 
one can obtain from 
(\ref{eq:introduction_08})
the short time and long time limits 
(\ref{eq:introduction_01})
and
(\ref{eq:introduction_02})
respectively. 
To the best of our knowledge, 
so far there still do not exist any derivations for the closed analytic expressions 
(\ref{eq:introduction_06})
or 
(\ref{eq:introduction_08})

Although the research on SFD can be traced back to 1960s, but until now there are not many studies that can provide a comprehensive 
description of the process. 
One notable exception is that of reference \cite{RodenbeckKargerHahn98}, 
which has derived the exact SFD propagator valid for all time scales based on the reflection principle of 
Chandrasekhar \cite{Chandrasekhar43}. 
Other attempts to model SFD include the early statistical and probabilistic models 
\cite{Fedders78,Karger92,Lin05,Harris65,Richards77,Levitt73,Beijeren83,Liggett85}
to the more recent ones based on fractional dynamics 
\cite{DemontisSuffritti06,Baqndyopadhyay08a,Baqndyopadhyay08b,TaloniLomholt08,LimTeo09}.
The main aims of this paper are two folds. 
First we introduce a new type of fractional generalized Langevin equation with external force to model SFD. 
Our second objective is to derive a new closed expression for the MSD of SFD. 
Despite that the fractional generalized 
Langevin equation under consideration does not lead to a closed expression for the MSD of SFD, 
it is still possible to show that it gives the correct short and long time limits under some specific conditions. 
It is possible to derive from various special cases of the fractional generalized Langevin equation
a completely new class of closed expressions for the MSD of SFD, which can be regarded as alternatives to
(\ref{eq:introduction_06}) and (\ref{eq:introduction_08}).
Our previous work 
\cite{LimTeo09} 
which though contains a detailed discussion of fractional generalized Langevin approach to SFD and its realization as a step fractional Brownian motion, 
it does not provide the derivation of closed expressions of the MSD for SFD. 
In addition, the present work also provides a discussion of the effective 
Fokker-Planck equation for SFD.

\section{Fractional Generalized Langevin Equation without External Force}
\label{sec:withoutExForce}
\noindent
In this section we want to study whether it is possible to model SFD using the fractional generalized Langevin equation without external force. 
For the purpose for subsequent discussion, 
let us first consider the following fractional generalized Langevin equation with external force: 
\begin{align}
  D^\alpha v(t) + \int_0^t \gamma(t-\tau)v(\tau)d\tau & = f(t) + \xi(t),
\label{eq:withoutExForce_001}
 \\
v(t) & = Dx(t),
\label{eq:withoutExForce_002}
\end{align}
where $0 < \alpha \leq 1$. 
$\xi(t)$ is an internal Gaussian noise with mean zero and covariance
\begin{align}
  \left<\xi(t)\xi(s)\right> & = C\!\left(\left|t-s\right|\right),
\label{eq:withoutExForce_003}
\end{align} 
$f(t)$  
is a time-dependent external force given by
\begin{align}
  f(t) & = a \frac{t^{-\kappa}}{\Gamma(1-\kappa)}, 
       \quad
       0 < \kappa \leq 1,
\label{eq:withoutExForce_004}
\end{align}
and
$\gamma(t)$
is the memory kernel which will be specified later on. 
The fractional derivative used in 
(\ref{eq:withoutExForce_001})
is the Caputo fractional derivative, which is defined by 
\cite{Podlubny99,MetzlerKlafter00,West03}
\begin{subequations}
\label{eq:withoutExForce_005}
\begin{align}
  D^\alpha g(t) &= I^{m-\alpha}D^m g(t),
\label{eq:withoutExForce_005a}
\end{align}
where the fractional integral
$I^\alpha$
is defined for
$\alpha > 0$
as
\begin{align}
  I^\alpha g(t) & = \frac{1}{\Gamma(\alpha)}\int_0^t (t - u)^{\alpha-1} g(u) du ,
\label{eq:withoutExForce_005b}
\end{align}
\end{subequations}
with 
$m-1 < \alpha \leq m$, 
and $m$ is positive integer.

Note that generalized Langevin equation 
(\ref{eq:withoutExForce_001})
with   
$f(t)=0$
has been studied by several authors 
\cite{Lutz01,Fa06,Fa07,LimTeo09}. 
The solution for the position process from 
(\ref{eq:withoutExForce_001}) and (\ref{eq:withoutExForce_002})
with initial conditions 
$x(0)=x_\circ$,
$v(0)=v_\circ$
is given by    
\begin{align}
  x(t) & = x_\circ + v_\circ I^{1-\alpha} G(t) + a I^{1-\kappa} G(t)
         + \int_0^t G(t-u) \xi(u) du,
\label{eq:withoutExForce_006}
\end{align}
where 
$G(t)$
is given by the inverse Laplace transform of
\begin{align}
  \tilde{G}(s) &= \frac{1}{s^{\alpha+1}+\tilde{\gamma}(s)s}.
\label{eq:withoutExForce_007}
\end{align}
The mean of $x(t)$ is given by
\begin{align}
  \bar{x} = \left<x(t)\right>
         & = x_\circ + \frac{v_\circ}{\Gamma(1-\alpha)}\int_0^t (t-u)^{-\alpha} G(u) du \nonumber \\
         & \quad + \frac{a}{\Gamma(1-\kappa)} \int_0^t (t-u)^{-\kappa} G(u) du
\label{eq:withoutExForce_008}
\end{align}
and its variance is 
\begin{align}
  \sigma^2 & = \left<\left(x(t) - \bar{x}\right)^2\right> \nonumber \\
           & = 2\int_0^t du G(u) \int_0^u dv C(u-v)G(v),
\label{eq:withoutExForce_009}
\end{align}
The MSD of $x(t)$ in terms of its mean and variance is
\begin{align}
  R^2 & = \left<\left(x(t) - x_\circ\right)^2\right>
        = \left(\bar{x}-x_\circ\right)^2 + \sigma^2.
\label{eq:withoutExForce_010}
\end{align}

Here we shall consider two cases without external force, 
that is with the constant
$a=0$; 
the case with external force will be studied in next section. 
The generalized Langevin equation
(\ref{eq:withoutExForce_001})
includes the following cases:

\vspace{0.5cm}
\noindent{\bf Case 1.}
$0<\alpha<1$,
$\gamma(t) = \lambda_1\delta(t)$,
$\lambda_1 > 0$,
$C(t) = c\delta(t)$,
and
$c > 0$

\noindent
(\ref{eq:withoutExForce_001})
becomes fractional Langevin equation and its solution has been considered by several authors 
\cite{KobolevRomanov00,LimMuniandy02,WestPicozzi02,LimEab06,LimLiTeo08}.
The variance and MSD of the position process at long time behaves like normal diffusion, 
that is
$\sigma^2 \sim t$ and $R^2 \sim t$ as $t \rightarrow \infty$.
On the other hand, we have
$\sigma^2 \sim t^{2\alpha+1}$,    
$R^2 \sim t^2$ as $t \rightarrow \infty$.
Note that if
(\ref{eq:withoutExForce_002})
is replaced by the fractional velocity 
$v(t)=D^\beta x(t)$, $0 < \beta < 1$, then for $x_\circ = 0$ one has
$R^2 \sim t^{2\beta-1}$
\cite{LimTeo09}.
This leads to SFD subdiffusion when
$\beta = 3/4$. 
For $t \rightarrow 0$, 
one obtains
$R^2 \sim t^{2(\alpha+\beta)-1}$, 
which gives normal diffusion if
$\beta = 1 - \alpha$, and ballistic motion if
$\alpha = \beta = 3/4$
\cite{LimTeo09}.

\vspace{0.5cm}
\noindent{\bf Case 2.}
$0<\alpha \leq 1$,
$\gamma(t) = \gamma_2(t)=\lambda_2\frac{t^{-\gamma}}{\Gamma(1-\gamma)}$,
and
$C(t) = c_\zeta t^{-\zeta}$,
$c_\zeta > 0$, 
$0 < \zeta \leq 1$

\noindent
This is the case of fractional generalized Langevin equation without external force and with power law type of memory kernel 
\cite{Lutz01,Fa06,Fa07,LimTeo09}.
The MSD corresponds to this case satisfy the following asymptotic properties. 
As $t \rightarrow 0$, 
one gets
\begin{align}
R^2  \sim 
\begin{cases}
  t^2, & \text{if} \ \alpha \geq \zeta/2 \\
  t^{2+2\alpha-\zeta}, & \text{if} \ \alpha < \zeta/2
\end{cases}.
\label{eq:withoutExForce_011}
\end{align}
In the case when 
$\alpha \geq \zeta/2$, 
the particle undergoes ballistic motion for very short times. 
However, 
when $\alpha > \zeta/2$, 
normal diffusion with 
$R^2 \sim t$  
is possible only if
$\zeta = 1+\alpha$ or $\zeta > 1$, 
which contradicts our assumption that
$\zeta \leq 1$.
For $t \rightarrow \infty$,
\begin{align}
  R^2 \sim 
  \begin{cases}
    t^{2\gamma-\zeta}, & \text{if} \ \gamma > \zeta/2 \\
    \log{t},           & \text{if} \ \gamma = \zeta/2  \\
    \text{constant},   & \text{if} \ \gamma < \zeta/2
  \end{cases},
\label{eq:withoutExForce_012}
\end{align}
which gives subdiffusion with $R^2 \sim \sqrt{t}$ when $\gamma = (2\zeta+1)/4$.
Here we note that the large time asymptotic behavior of MSD is independent of the fractional order $\alpha$
of the generalized fractional Langevin equation. 
From (\ref{eq:withoutExForce_011}) one notices that the particle does not diffuse normally for short times, 
though it can undergo subdiffusion for $\gamma > \zeta/2$ after sufficiently long time.
However, if the particle is assumed to undergo ballistic motion at very short times, 
then the generalized Langevin equation in this case gives the correct short and long time limits of the MSD for SFD.

\section{Fractional Generalized Langevin Equation with External Force}
\label{sec:withExForce}
\noindent
In this section we consider the generalized Langevin equation 
(\ref{eq:withoutExForce_001})
with external force under the following general setting:

\vspace{0.5cm}
\noindent{\bf Case 3.}
$0 < \alpha < 1$,
$\gamma(t) = \lambda_1 \delta(t) + \lambda_2 \frac{t^{-\gamma}}{\Gamma(1-\gamma)}$
and
$a \neq 0$.

\noindent
A special case with
$\alpha = 1$,
$\gamma = 1/2$ and 
$a = 0$
has been considered recently in 
\cite{TaloniLomholt08}. 
Solution to the position process is then given by 
(\ref{eq:withoutExForce_006}) 
with   
$G(t)$ 
is given by the inverse Laplace transform of
\begin{align}
  \tilde{G}(s) & = \frac{1}{s^{\alpha+1}+ \lambda_1 s + \lambda_2 s^\gamma}.
\label{eq:withExForce_001}
\end{align}
In order to study the asymptotic properties of the solution, 
we expand
$\tilde{G}(s)$
in the following series form
\begin{align}
  \tilde{G}(s) & = \frac{s^{-\gamma}}{s^{\alpha+1-\gamma} + \lambda_2}
                   \sum_{n=0}^\infty \left[\frac{s^{1-\gamma}}{s^{\alpha+1-\gamma} + \lambda_2}\right]^n.
\label{eq:withExForce_002}
\end{align}
One can express the solution in terms of 
the two parameter Mittag-Leffler function
$E_{\alpha,\beta}(z)$
by using the following Laplace transform relation:
\begin{align}
  L\left[t^\beta E_{\alpha,\beta}\left(\lambda t^\alpha\right)\right]
      & = \frac{s^{\alpha-\beta}}{s^\alpha - \lambda},
\label{eq:withExForce_003}
\end{align}
where
$E_{\alpha,\beta}(z)$
is defined by 
\cite{Erdelyi3_53}
\begin{align}
  E_{\alpha,\beta}(z) & = \sum_{n=0}^\infty \frac{z^n}{\Gamma(\alpha n + \beta)}.
\label{eq:withExForce_004}
\end{align}
Now define 
\begin{subequations} 
\label{eq:withExForce_005}
\begin{align} 
G_\circ(t) & = t^\alpha E_{\alpha-\gamma+1,\alpha+1}\left(-\lambda_2 t^{\alpha-\gamma+1}\right),
\label{eq:withExForce_005a}
\\
G_\circ(t)^* & = t^{\alpha -1} E_{\alpha-\gamma+1,\alpha}\left(-\lambda_2 t^{\alpha-\gamma+1}\right),
\label{eq:withExForce_005b}
\end{align}
\end{subequations}
and
\begin{align}
  G_\circ^{*n}(t) & = \int_0^t du_1 G_\circ^*\left(t-u_1\right)
                      \int_0^{u_1} du_2 G_\circ^*\left(u_1 - u_2\right)
                      \cdots \nonumber \\
                  & \quad 
                      \cdots
                      \int_0^{u_{n-2}} du_{n-1} G_\circ^*\left(u_{n-2} - u_{n-1}\right)
                      G_\circ^*\left(u_{n-1}\right)
\label{eq:withExForce_006}
\end{align}
for $n \geq 2$. 
We then have
\begin{align}
  G(t) & = G_\circ (t) + \sum_{n=0}^\infty \left(-\lambda_1\right)^n
                         \int_0^t du G_\circ(t - u)G_\circ^{*n}(u)
\label{eq:withExForce_007}
\end{align}
Some properties of
$G(t)$
which are necessary for obtaining the asymptotic limits of MSD are given in 
\ref{sec:appendixA}.

We shall also need the following asymptotic expansion of 
Mittag-Leffler function 
\cite{Erdelyi3_53}
to obtain the asymptotic properties of the variance and MSD. 
For $z \rightarrow \infty$,
\begin{align}
  E_{\alpha,\beta}(-z) & = - \sum_{n=1}^N \frac{(-1)^{n-1}z^{-n}}{\Gamma(\beta - n\alpha)}
                             + \mathcal{O}\left(|z|^{-1-N}\right), \nonumber \\
                     & \hspace{2cm} \left|\arg(z)\right| < \left(1 - \frac{\alpha}{2}\right)\pi
\label{eq:withExForce_008}
\end{align}
and for $z \rightarrow 0$,
\begin{align}
E_{\mu,\nu}(-z) & \sim \frac{1}{\Gamma(\nu)} + \mathcal{O}(z),
\label{eq:withExForce_009}
\end{align}

\vspace{0.5cm}
\noindent
{(i). Short time limit}

For $t \rightarrow 0$, we have 
\begin{align}
  G(t) & \underset{t \to 0}{\sim} \frac{t^\alpha}{\Gamma(\alpha+1}
          - \lambda_2 \frac{t^{2\alpha-\gamma+1}}{\Gamma(2\alpha-\gamma+2)} 
          - \lambda_1 \frac{t^{2\alpha}}{\Gamma(2\alpha+1)}.
\label{eq:withExForce_010}
\end{align}
Since $\gamma \leq  1$, one gets
\begin{align}
    G(t) & \underset{t \to 0}{\sim} \frac{t^\alpha}{\Gamma(\alpha+1}
          - \lambda_1 \frac{t^{2\alpha}}{\Gamma(2\alpha+1)}.
\label{eq:withExForce_011}
\end{align}
From (\ref{eq:withoutExForce_006}) 
and (\ref{eq:appendixA_001}) in \ref{sec:appendixA}
one gets
\begin{align}
  \bar{x} - x_\circ &= v_\circ I^{1-\alpha} G(t) + a I^{1-\kappa} G(t) 
\label{eq:withExForce_012}
 \\
                    &\underset{t \to 0}{\sim} 
                      v_\circ\left[t - \lambda_1 \frac{t^{\alpha+1}}{\Gamma(\alpha+2)}\right] 
                     \nonumber \\
                    & \quad + a\left[
                      \frac{t^{\alpha-\kappa+1}}{\Gamma(\alpha-\kappa+2)}
                      -\lambda_1\frac{t^{2\alpha-\kappa+1}}{\Gamma(2\alpha-\kappa+2)}
                      \right].
\label{eq:withExForce_013}
\end{align}
We therefore obtain
\begin{subequations}
\label{eq:withExForce_014}
\begin{align}
  \bar{x} - x_\circ &\underset{t \to 0}{\sim} 
                      v_\circ t + a\frac{t^{\alpha-\kappa+1}}{\Gamma(\alpha-\kappa+2)},
\label{eq:withExForce_014a}
\end{align}
and
\begin{align}
  \bar{x} - x_\circ &\underset{t \to 0}{\sim}
  \begin{cases}
    v_\circ t,                  & a = 0, \text{or} \ \alpha > \kappa \\
    \left(v_\circ + a\right)t,  & a \neq 0, \alpha = \kappa         \\
    a\frac{t^{\alpha-\kappa+1}}{\Gamma(\alpha-\kappa+2)}, & a \neq 0, \alpha < \kappa       
  \end{cases}.
\label{eq:withExForce_014b}
\end{align}
\end{subequations}
Using the following expression for variance
(see \ref{sec:appendixB} for its derivation)
\begin{align}
  \sigma^2 &= 2k_B T\left\{\int_0^t du G(u) - \int_0^t du G(u)D^\alpha G(u)\right\},
\label{eq:withExForce_015}
\end{align}
one gets
\begin{align}
  \sigma^2 &\underset{t \to 0}{\sim} 2k_B T 
            \lambda_1 \left[\frac{t^{2\alpha+1}}{(2\alpha+1)\Gamma^2(\alpha+1)}\right].
\label{eq:withExForce_016}
\end{align}
Here we remark that in the derivation of (\ref{eq:withExForce_015}), 
we have made use of the fluctuation-dissipation theorem 
(see \ref{sec:appendixB}). 
It has been pointed out that the FD theorem fails in the presence of external random noise 
\cite{Kubo66}. 
However, the external force in the fractional generalized Langevin equation 
(\ref{eq:withoutExForce_001})
is a non-random force, 
so FD theorem is applicable in this case up to first order of the external force term 
\cite{WangTokuyama99}. 
Note that we also assume that the FD theorem applies to the fractional generalized Langevin equation, 
as done in references 
\cite{TaloniLomholt08,LimTeo09b}.
Usual fluctuation-dissipation theorem is valid for SFD system provided the fluid or gas is elastic and diluted 
\cite{Villamaina08}. 
In the case of strongly inelastic and dense systems, 
fluctuation-dissipation formula fails and a more general form of fluctuation-dissipation relation has to be used.

By noting that the short time behavior of variance is of the order $t^{2\alpha+1}$, 
therefore for $\alpha > 1/2$, 
it is larger than   
$\left(\bar{x} - x_\circ\right)^2$
given by the square of 
(\ref{eq:withExForce_014b}), 
thus it is only this term that dominates the short time limit of MSD. 
This gives
\begin{subequations}
\label{eq:withExForce_017}
\begin{align}
  R^2 &\underset{t \to 0}{\sim}
  \begin{cases}
    v_\circ^2 t^2,                  & a = 0 \ \text{or} \ \alpha > \kappa \\
    \left(v_\circ + a\right)^2t^2,  & a \neq 0, \alpha = \kappa         \\
    a^2\frac{t^{2\alpha-2\kappa+2}}{\Gamma^2(\alpha-\kappa+2)}, & a \neq 0, \alpha < \kappa       
  \end{cases}.
\label{eq:withExForce_017a}
\end{align}
The short time limits associated with $\alpha > 1/2$ given by (\ref{eq:withExForce_017a})
implies that the two cases with  $\alpha > \kappa$ (zero external force) 
and $\alpha = \kappa$ (non-zero external force) are both ballistic in nature. 
The case corresponds to non-zero external force with $\alpha < \kappa$ is non-ballistic, 
and it leads to normal diffusion if $\kappa = \alpha +1/2$.  

For $\alpha = 1/2$, one gets
\begin{align}
  R^2 &\underset{t \to 0}{\sim}
  \begin{cases}
    \left(v_\circ^2 + \frac{4k_BT\lambda_1}{\pi}\right)t^2,                  & a = 0 \ \text{or} \ 1/2 > \kappa \\
    \left[\left(v_\circ + a\right)^2 + \frac{4k_BT\lambda_1}{\pi}\right]t^2,  & a \neq 0, \kappa = 1/2         \\
    a^2\frac{t^{3-2\kappa}}{\Gamma^2(3/2-\kappa)}, & a \neq 0, 1/2 < \kappa       
  \end{cases}.
\label{eq:withExForce_017b}
\end{align}
The short time limits given by (\ref{eq:withExForce_017b}) are quite similar to those in (\ref{eq:withExForce_017a}). 
Again, the first two cases with $\alpha =1/2 > \kappa$ (zero external force) 
and $\alpha =1/2 = \kappa$ (non-zero external force) both lead to ballistic motion, 
while the third case with $\alpha =1/2 < \kappa$ (zero external force) is non-ballistic, 
it gives normal diffusion only if $\kappa = 1$.

Finally, for $\alpha < 1/2$ one obtains
\begin{align}
  R^2 &\underset{t \to 0}{\sim}
  \begin{cases}
    2k_B T\lambda_1 \left[\frac{t^{2\alpha+1}}{(2\alpha+1)\Gamma^2(\alpha+1)}\right],   & a = 0 \ \text{or} \ \alpha > \kappa \\
    a^2\frac{t^{2\alpha-2\kappa+2}}{\Gamma^2(\alpha-\kappa+2)}, & a \neq 0, \alpha < \kappa       
  \end{cases}.
\label{eq:withExForce_017c}
\end{align}
\end{subequations}
Both cases in (\ref{eq:withExForce_017c}) can not be ballistic. 
The first case with $\alpha > \kappa$   
or zero external force gives superdiffusion, 
hence does not describe SFD. 
For non-zero external force with $\alpha < \kappa$, 
the short time limit leads to normal diffusion if $\kappa = \alpha + 1/2$, 
which is the same condition as for the third case in (\ref{eq:withExForce_017a}) and (\ref{eq:withExForce_017b}). 
In other words, we always obtain normal diffusion in the short time limit for $\kappa = \alpha + 1/2$.

\vspace{0.5cm}
\noindent
{(ii). Long time limit}

For $t \gg 1$, we have from (\ref{eq:withExForce_013})
\begin{align}
  \bar{x} - x_\circ  & \underset{t \to \infty}{\sim}
                         v_\circ\left[\frac{t^{\gamma - \alpha}}{\lambda_2\Gamma(\gamma - \alpha +1)}\right]
                       + a\left[\frac{t^{\gamma - \kappa}}{\lambda_2\Gamma(\gamma - \kappa +1)}\right].
\label{eq:withExForce_018}
\end{align}
When $\kappa > \alpha$ (\ref{eq:withExForce_018}) becomes
\begin{align}
  \bar{x} - x_\circ  & \underset{t \to \infty}{\sim}
                         v_\circ\left[\frac{t^{\gamma - \alpha}}{\lambda_2\Gamma(\gamma - \alpha +1)}\right].
\label{eq:withExForce_019}
\end{align}
From the variance relation (\ref{eq:withExForce_015}) one gets
\begin{align}
  \sigma^2 & \underset{t \to \infty}{\sim}
              2k_B T \left\{
                       \frac{t^\gamma}{\lambda_2\Gamma(\gamma+1)}
                     \right. \nonumber \\
          & \qquad\qquad  - \left.
                       \frac{t^{2\gamma - \alpha -1}}{\lambda_2^2(2\gamma - \alpha -1)\Gamma(\gamma)\Gamma(\gamma - \alpha)}
                     \right\} \nonumber\\
          & \underset{t \to \infty}{\sim} 2k_B T \frac{t^\gamma}{\lambda_2\Gamma(\gamma+1)},
\label{eq:withExForce_020}
\end{align}
since $\gamma < \alpha +1$.
Using (\ref{eq:withoutExForce_010}) for the MSD and together with (\ref{eq:withExForce_019})
and (\ref{eq:withExForce_020}), 
we obtain the following three cases for the long time limit for MSD:

\begin{subequations}
\label{eq:withExForce_021}
\noindent
If $\gamma < 2\alpha$,
\begin{align}
  R^2 & \underset{t \to \infty}{\sim} 2k_B T \frac{t^\gamma}{\lambda_2\Gamma(\gamma+1)}.
\label{eq:withExForce_021a}
\end{align}
If $\gamma = 2\alpha$,
\begin{align}
  R^2 & \underset{t \to \infty}{\sim} \left[
                                           \frac{2k_B T}{\lambda_2\Gamma(\gamma+1)}
                                          + \frac{v_\circ^2}{\lambda_2^2\Gamma^2\left(\frac{1}{2}\gamma + 1\right)}
                                      \right]t^\gamma.
\label{eq:withExForce_021b}
\end{align}
If $\gamma > 2\alpha$, then we have
\begin{align}
  R^2 & \underset{t \to \infty}{\sim} v_\circ^2\left[
                                          \frac{t^{2(\gamma-\alpha)}}{\lambda_2^2\Gamma^2\left(\gamma - \alpha + 1\right)}
                                      \right].
\label{eq:withExForce_021c}
\end{align}
\end{subequations}
We remark that the initial velocity can be determined by the equipartition principle of 
kinetic energy $v_\circ^2 = k_BT$. 
One may therefore assume that coefficient $a$ for the external force is proportional 
to $\sqrt{k_BT}$ and takes the form $a = a_1\sqrt{k_BT}$, $a_1$ is a positive constant. 
Note that the fractional order of the Langevin equation 
or $\alpha$ does not appear in the long time limits of MSD for the position process 
in the case of $\gamma < 2\alpha$ and the case $\gamma = 2\alpha$, just like for
Case 1 and Case 2 (without external force) discussed in Section \ref{sec:withoutExForce}. 
However, when $\gamma > 2\alpha$, 
the long time limit of MSD $R^2$ given by (\ref{eq:withExForce_021c}) is dependent on $\alpha$. 
On the other hand, the exponent of the external force $\kappa$ is absent in (\ref{eq:withExForce_021}).

Here we would like take note of the importance of initial conditions on SFD 
which has been pointed out in a recent work
\cite{BarkaiSilbey09}.
Here we recall that according to equations 
(\ref{eq:withExForce_016}) and (\ref{eq:withExForce_017}), 
the short time limits for the variance and MSD have different time dependence. 
If now we suppose the initial condition is $x(0)=0$ instead of $x(0) = x_\circ$, 
and we assume the process is centred, that is $\bar{x} = 0$, then we have $R^2=\sigma^2$. 
Now the short time limit of MSD has the same time dependence as the variance. 
For the long time limits, both the variance and MSD 
have the same time dependence except for 
$\gamma > 2\alpha$, 
and they are independent of $x(0)$. 
A similar observation had been obtained in reference 
\cite{LimTeo09}

The above discussion shows that 
it is possible to obtain the correct short and long time behaviors of the SFD using the generalized fractional Langevin
equation (\ref{eq:withoutExForce_001})  and (\ref{eq:withoutExForce_002}) under certain specific conditions.
In the following we want to consider some limiting and special cases.

\vspace{0.5cm}
\noindent{\bf Case 3a.}
Overdamped case:

\noindent
Consider the overdamped case such that the Newton acceleration or ballistic term 
(for $\alpha = 1$) 
or the fractional acceleration term 
(for $\alpha \neq 1$) 
is neglected. 
Now the second term in the expression for variance (\ref{eq:withExForce_015}) will be absent, 
and the mean is simply
\begin{align}
  \left<x_1(t)\right> &= x_\circ,
\label{eq:withExForce_022}
\end{align}
such that
\begin{align}
 R^2 = \sigma^2 = 2k_BT \int_0^t du G(u).
 \label{eq:withExForce_023}
\end{align}

Let us calculate the overdamped case in more detail. 
Now the Laplace transform of $G(t)$ is
\begin{align}
  \tilde{G}(s) &= \frac{1}{\lambda_1 s + \lambda_2 s^\gamma}
                = \zeta \frac{s^{(1-\gamma)-1}}{s^{1-g} + \lambda},
 \label{eq:withExForce_024}
\end{align}
where $\zeta = 1/\lambda_1$ and $\lambda = \lambda_2/\lambda_1$, thus
\begin{align}
  G(t) & = \zeta E_{1-\gamma,1}\left(-\lambda t^{1-\gamma}\right).
 \label{eq:withExForce_025}
\end{align}
One now gets
\begin{align}
  R^2 = \sigma^2 & = 2k_BT\zeta \int_0^\infty d\tau E_{1-\gamma,1}\left(-\lambda t^{1-\gamma}\right) \nonumber \\
     & = 2k_BT\zeta E_{1-\gamma,2}\left(-\lambda t^{1-\gamma}\right).
 \label{eq:withExForce_026}
\end{align}
The asymptotic behaviors of  MSD are given by
\begin{align}
  R^2 & \sim
  \begin{cases}
    2k_BT\zeta t, &  t \rightarrow 0  \\
    2k_BT\zeta \frac{t^\gamma}{\lambda\Gamma(1-\gamma)} & t \rightarrow \infty
  \end{cases}.
 \label{eq:withExForce_027}
\end{align}
Therefore for $\gamma = 1/2$, one gets the correct subdiffusion behavior corresponds to the SFD.

\vspace{0.5cm}
\noindent{\bf Case 3b.}
$0 < \alpha <1$,
$\gamma(t) = \lambda_2 \frac{t^{-\gamma}}{\Gamma(1-\gamma)}$
and
$a \neq 0$

\noindent
This is the case of (\ref{eq:withoutExForce_001})
with external force and single term of memory kernel 
(with $\lambda_1 = 0$).
We have   
$
G(t) =  G_\circ(t) =  t^\alpha E_{\alpha -\gamma -1, \alpha + 1}
\left(-\lambda_2 t^{\alpha - \gamma +1}\right)
$
from (\ref{eq:withExForce_005a}) as the Green function. 
Using  (\ref{eq:appendixA_002}) and (\ref{eq:appendixA_003}) in \ref{sec:appendixA}, 
together with the asymptotic formulas of Mittag-Leffler functions (\ref{eq:withExForce_008}) and (\ref{eq:withExForce_009}), 
one gets from equations in (\ref{eq:withExForce_015}) the asymptotic properties of the variance. 
For $t \rightarrow \infty$
\begin{align}
  \sigma^2 & \sim 2k_BT\left[
                       \frac{t^\gamma}{\lambda_2\Gamma(\gamma+1)}
                      -\frac{t^{2\gamma - 1 -\alpha}}
                            {\lambda_2^2(2\gamma -1 -\alpha)
                             \Gamma(\gamma)\Gamma(\gamma-\alpha)}
                     \right] \nonumber \\
           & \sim 2k_BT\frac{t^\gamma}{\lambda_2\Gamma(\gamma+1)},
 \label{eq:withExForce_028}
\end{align}
where we use the fact that $\gamma < 1 + \alpha$. 
On the other hand, for $t \rightarrow 0$
\begin{align}
    \sigma^2 & \sim 2k_BT\left[\lambda_2
                               \frac{t^{2\alpha - \gamma +2}}
                                    {(2\alpha - \gamma +2)
                                     \Gamma(\alpha+1)\Gamma(\alpha - \gamma+2)}
                     \right].
 \label{eq:withExForce_029}
\end{align}

Now consider the mean
\begin{subequations}
 \label{eq:withExForce_030}
\begin{align}
  \bar{x} - x_\circ &= v_\circ tE_{\alpha - \gamma +1,2}\left(-\lambda_2t^{\alpha - \gamma +1}\right) 
\nonumber 
 \label{eq:withExForce_030a}
\\        
                    & \quad + at^{\alpha-\kappa+1}E_{\alpha - \gamma +1,\alpha - \kappa+2}\left(-\lambda_2t^{\alpha - \gamma +1}\right) \\
                    & \underset{t \to \infty }\sim v_\circ \frac{t^{\gamma -\alpha}}{\lambda_2\Gamma(\gamma - \alpha +1)}
                      + a \frac{t^{\gamma - \kappa}}{\lambda_2\Gamma(\gamma - \kappa +1)},
 \label{eq:withExForce_030b}
\end{align}
\end{subequations}
which decays to zero for 
$\alpha > \gamma$ and $\kappa > \gamma$ as $t \rightarrow \infty$. 
The MSD behaves asymptotically in the same way as the variance, i.e.
\begin{align}
  R^2 & \sim 2k_BT \frac{t^\gamma}{\lambda_2\Gamma(\gamma + 1)}, 
        \quad t \rightarrow \infty.
 \label{eq:withExForce_031}
\end{align}

One gets from (\ref{eq:withExForce_030a}) the small time limit
\begin{align}
  \bar{x} - x_\circ & \sim v_\circ t + a \frac{t^{\alpha - \kappa +1}}{\Gamma(\alpha - \kappa +2)},
                      \quad  t\rightarrow 0.
 \label{eq:withExForce_032}
\end{align}
Since $0 < \alpha \leq 1$ and $0 < \kappa \leq 1$, if $\alpha < \kappa$, we have
\begin{align}
  \bar{x} - x_\circ & \sim a \frac{t^{\alpha - \kappa +1}}{\Gamma(\alpha - \kappa +2)},
                              \quad  t\rightarrow 0.
 \label{eq:withExForce_033}
\end{align}
We can now conclude that the MSD takes the form
\begin{align}
  R^2 & \sim a^2 \frac{t^{2\alpha - 2\kappa +2}}{\Gamma^2(\alpha - \kappa +2)}
\nonumber \\
      & \quad + 2k_BT\left[\lambda_2
                       \frac{t^{2\alpha - \gamma +2}}
                       {(2\alpha - \gamma +2)
                         \Gamma(\alpha+1)\Gamma(\alpha - \gamma+2)}
                     \right].
 \label{eq:withExForce_034}
\end{align}
The case $\alpha = 1/2$, $\gamma = 1/2$, $\kappa = 1$ gives the MSD as 
\begin{align}
  R^2 & \sim 
  \begin{cases}
    a^2 \frac{t}{\Gamma^2(3/2)} &  t \rightarrow 0  \\
    2k_BT\frac{t^{1/2}}{\lambda_2\Gamma(3/2)} &  t \rightarrow \infty
  \end{cases}
 \label{eq:withExForce_035}
\end{align}
which satisfies the behavior of SFD.

As a manifestation of the non-passing many-particle syatem, a realistic physical model of SFD should include the long-range inter-particle correlation, and hence the particle density term.
However, in our model the correlation function does not have a closed analytic form, 
we do not pursue it since SFD can be characterized by its MSD, which can be calculated more easily.
The particle density and others many-body effect have been absorbed into the memory kernel term and the random force,
and it can be linked to the variance and MSD via the diffusion coefficient 
$D$ and SFD mobility $F$ (see for example 
\cite{TaloniLomholt08}). 
We shall identify in next section the constants 
in the memory kernel with the diffusion coefficient and SFD mobility. 
The long-range correlation property of SFD in our model is manifested in the realization of the single-file subdiffusion process as fractional Brownian motion with 
Hurst index $H = 1/4$. 
The long-range dependence for such a realization can be verified by using the argument given in 
\cite{LimMuniandy03}.

We remark that the mathematical modeling of SFD based on its MSD will not be unique. 
Recall that a Gaussian random process is determined (up to a multiplicative constant) by its mean and covariance function. 
Two different Gaussian processes can have the same MSD or variance 
but with different covariance functions. 
Thus, any modeling based on fractional Langevin equations that give the correct short and long time limits to 
the MSD of SFD can at best be regarded as one of the possible models. 
In order to have a more realistic and concrete model of SFD, 
it is necessary to take into account the physical interactions and interplays between the boundary conditions and the particle motion.


\section{New Closed Expressions for MSD of SFD}
\label{sec:clossExp}
\noindent
From the above discussion we see that it is not possible to obtain a close analytic expression for the MSD 
for the position process from the solution to 
(\ref{eq:withoutExForce_001})
and 
(\ref{eq:withoutExForce_002}), 
even though we are able to show that both the long time and short time asymptotic properties of the MSD agree with that for SFD. 
However, for the overdamped case one has the closed expression (\ref{eq:withExForce_026}) for the MSD.
By letting $\gamma = 1/2$, (\ref{eq:withExForce_026}) gives the correct short and long time limits for the MSD of SFD.

Let us consider this special case in more detail.
If we let $\gamma = 1/2$, and with appropriate values for $\zeta$  and $\lambda$, 
we can then show that   
$R^2 = 2k_BT\zeta tE_{1/2,2}\left(-\lambda t^{1/2}\right)$
can provide the correct description of SFD.  
By comparing this expression with (\ref{eq:introduction_03}) one gets in the limit $t \rightarrow 0$, 
\begin{align}
  R^2 & \sim 2k_BT\zeta t = l^2(1 -\theta) \frac{t}{\tau},
\label{eq:clossExp_001}
\end{align}
which requires $\zeta = l^2(1-\theta)/(2k_BT\tau)$. 
On the other hand, by comparing it with (\ref{eq:introduction_04})
one has for $t \rightarrow \infty$,
\begin{gather}
  R^2 \sim 2k_BT\zeta t \frac{\left(\lambda t^{1/2}\right)^{-1}}{\Gamma(3/2)}
      = \frac{4k_BT\zeta}{\lambda\sqrt{\pi}}t^{1/2} \nonumber \\
      = l^2 \frac{1-\theta}{\theta}\sqrt{\frac{2}{\pi}}\sqrt{\frac{t}{\tau}},
\label{eq:clossExp_002}
\end{gather}
which requires $\lambda =\theta\sqrt{2/\tau}$. 
Therefore, the expression $R^2 = 2k_BT\zeta tE_{1/2,2}\left(-\lambda t^{1/2}\right)$
with $\zeta = l^2(1-\theta)/(2k_BT\tau)$ and $\lambda = \theta\sqrt{2/\tau}$
provides a new alternative expression for the MSD of SFD.

\begin{figure*}[h!]
  \centering
    \subfloat[]{
    \includegraphics*[viewport=30 30 680 480,scale=0.35]{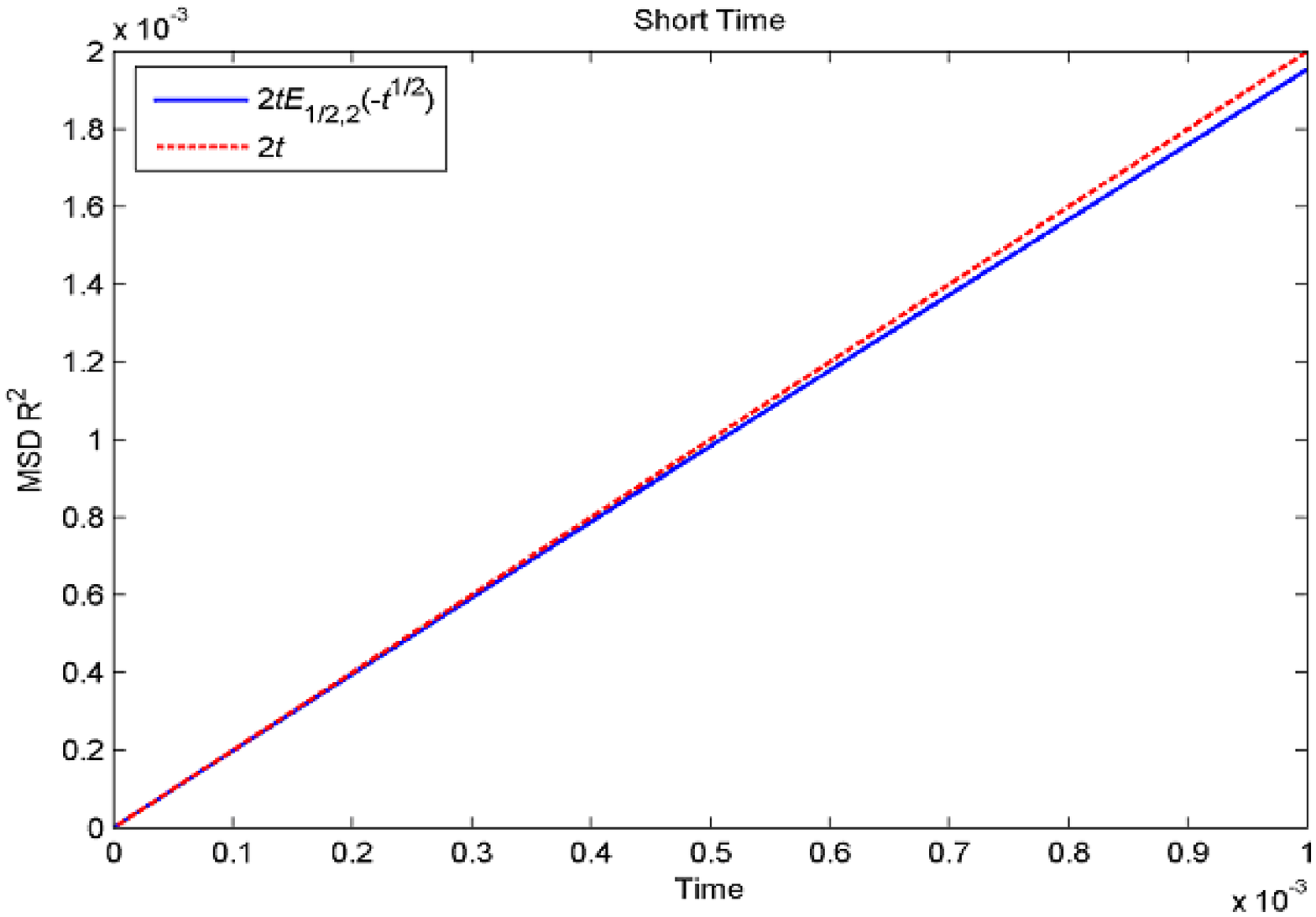}}
    \subfloat[]{
    \includegraphics*[viewport=30 30 680 480,scale=0.35]{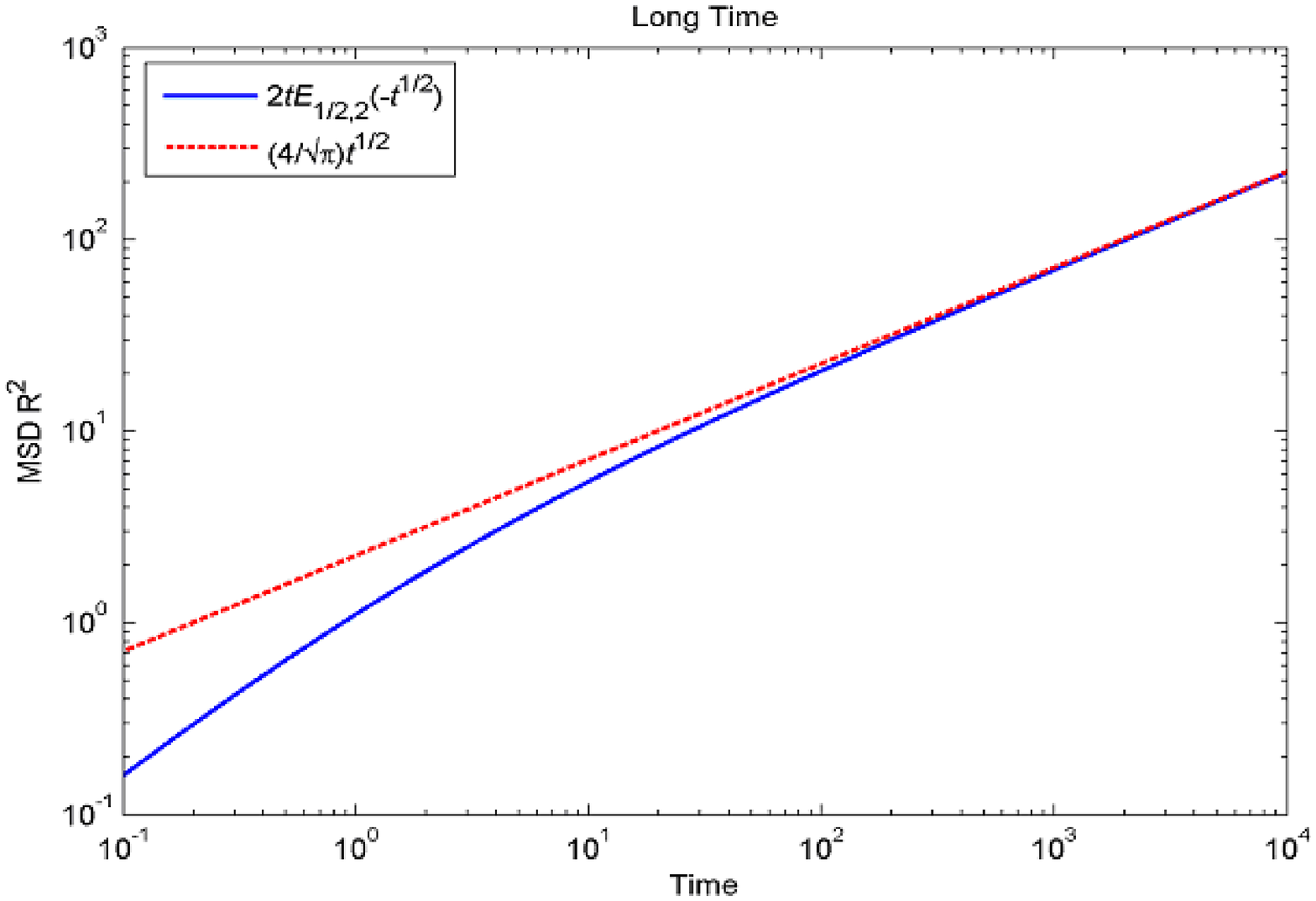}}
 
  \caption{Short- and long-time simulations for MSD $2tE_{1/2,2}\left(-t^{1/2}\right)$.}
  \label{fig:SFD01}
\end{figure*}

\begin{figure*}[h!]
  \centering
     \subfloat[]{
    \includegraphics*[viewport=30 30 680 480,scale=0.35]{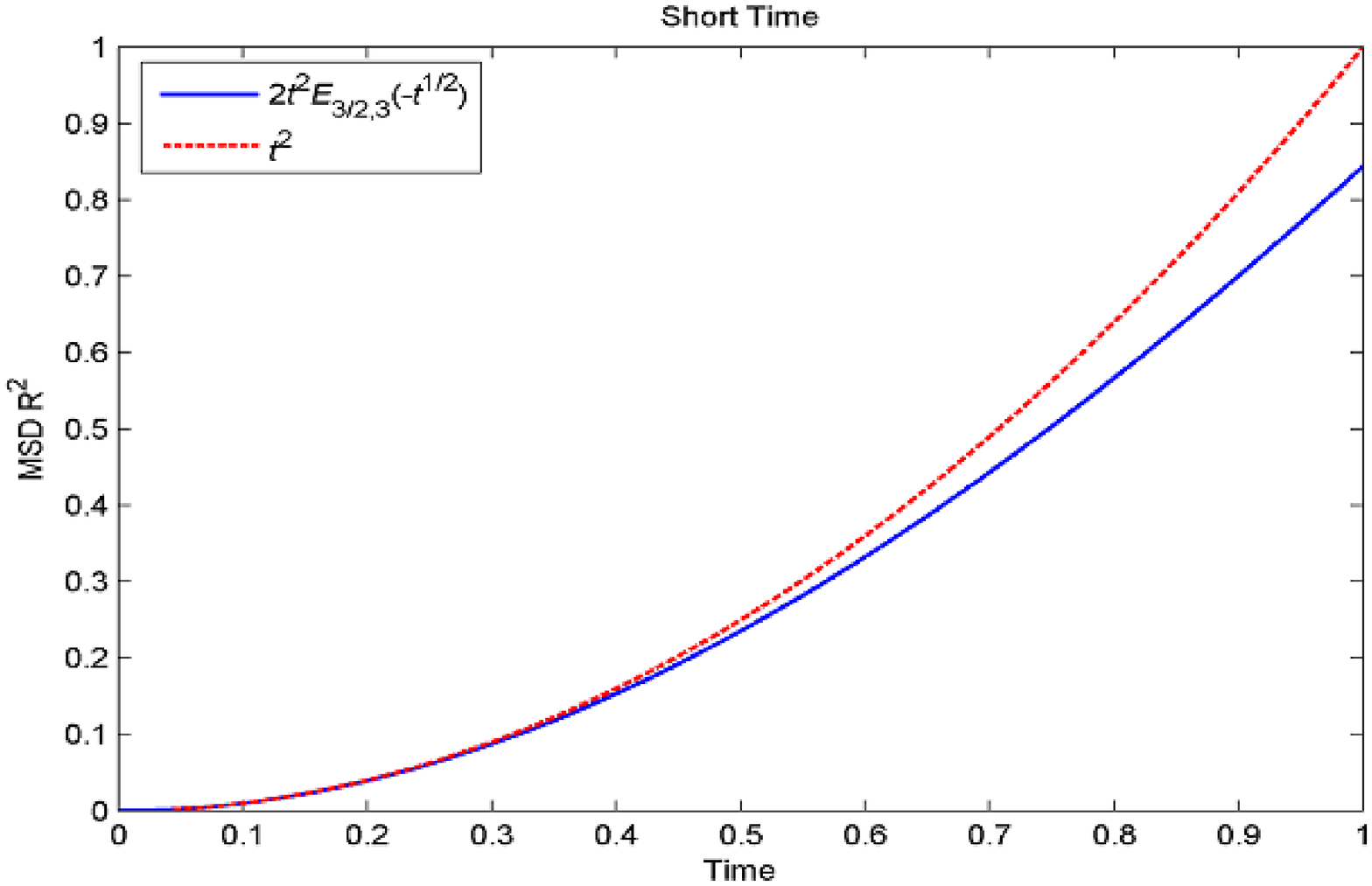}}
    \subfloat[]{
    \includegraphics*[viewport=30 30 680 480,scale=0.35]{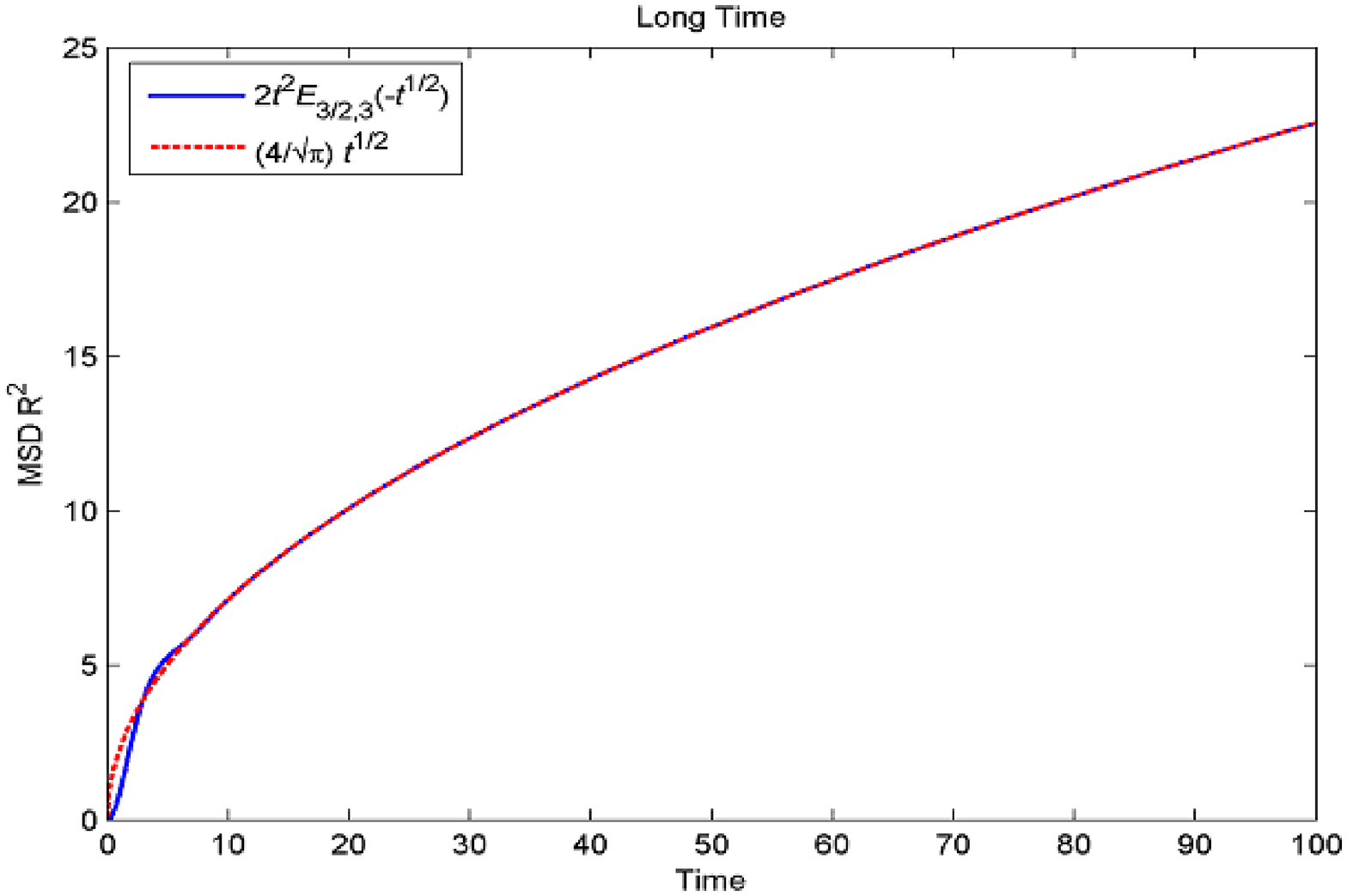}}
 
  \caption{Short- and long-time simulations for MSD $2t^2E_{3/2,3}\left(-t^{3/2}\right)$.}
  \label{fig:SFD02}
\end{figure*}

By noting that both $E_{1/2,2}(t)$ and $E_{1/2,\beta}(t)$, $\beta > 0$ 
have the same short and long time limits (up to a multiplicative constant), it is then possible to generalized the above closed
expression for the MSD to the following more general expression
\begin{align}
  R^2 & = 2k_BT\zeta^\prime t E_{1/2,\beta}\left(-\lambda^\prime t^{1/2}\right),
\label{eq:clossExp_003}
\end{align}
for $\beta > 0$. 
Again, by comparing (\ref{eq:clossExp_003}) with (\ref{eq:introduction_03}) and (\ref{eq:introduction_04}) one gets 
$\zeta^\prime = l^2(1- \theta)\Gamma(\beta)/(2k_BT\tau)$  
and  $\lambda^\prime = \theta\Gamma(\beta-\frac{1}{2})/\sqrt{2\tau}$.
Expression (\ref{eq:clossExp_003}) gives the family of closed form expressions for SFD, 
and they can be regarded as alternatives to the equation (\ref{eq:clossExp_003}) obtained by Brandani \cite{Brandani96}. 
Simulations of  
$R^2 = 2\zeta tE_{1/2,2}\left(-\lambda t^{1/2}\right)$ 
and the more general case   
$R^2 = 2\zeta tE_{1/2,\beta}\left(-\lambda t^{1/2}\right)$ 
with some specific values of $\zeta$, $\lambda$ and $\beta$ are given in Figures \ref{fig:SFD01} and \ref{fig:SFD02}.

\begin{figure*}[h!]
  \centering
  \includegraphics*[viewport=30 30 680 480,scale=0.35]{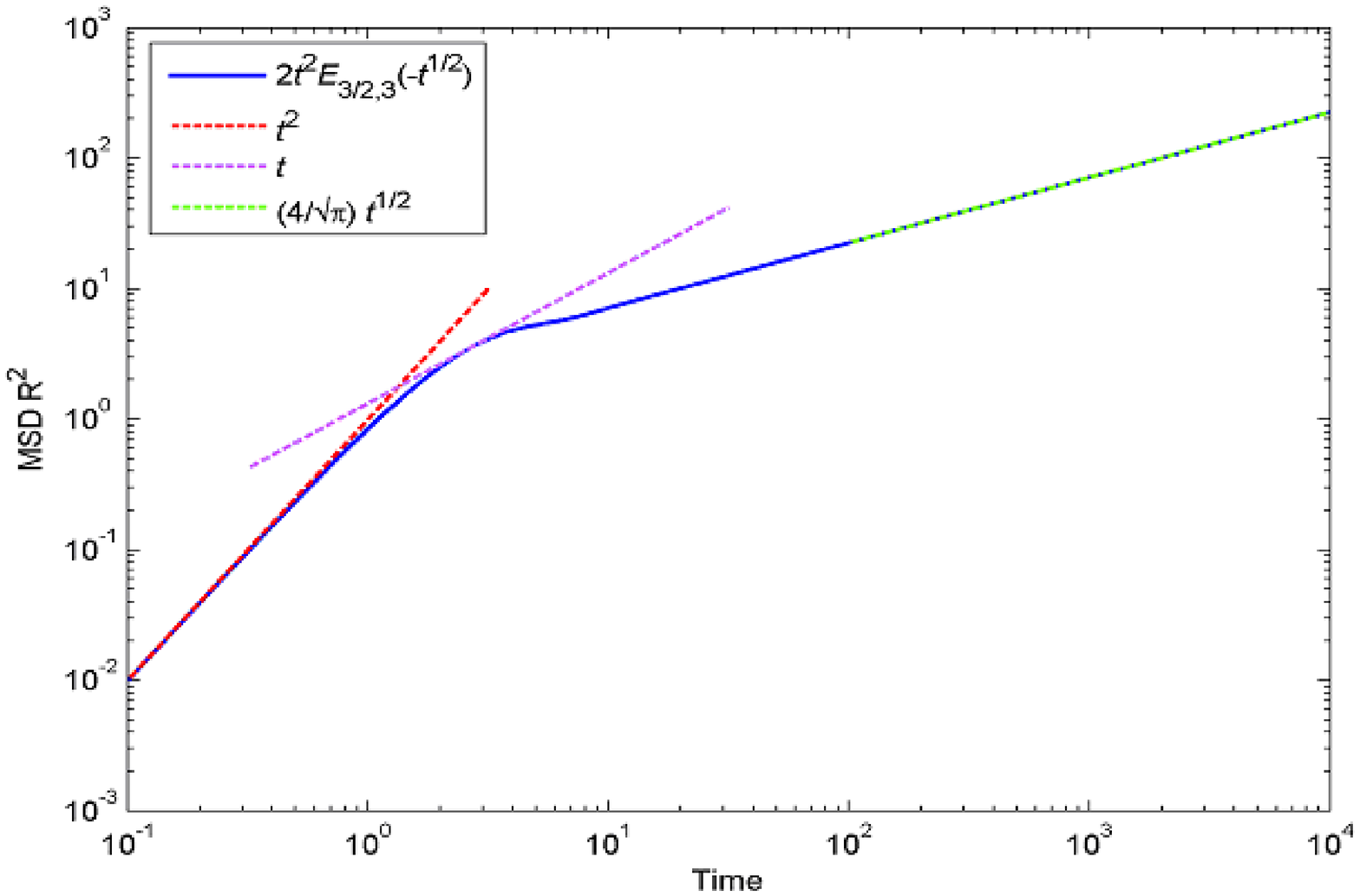}
  \caption{Plots showing the three regimes of SFD according to MSD $R^2 = 2t^2E_{3/2,3}\left(-t^{3/2}\right)$.}
  \label{fig:SFD03}
\end{figure*}

The SFD model having transition from normal diffusion to subdiffusion regime can only be regarded as an approximation to the real SFD system. 
At very short times such a system first undergoes ballistic motion with 
${\rm MSD} \sim t^2$.
Ballistic motion occurs before the particle has had a chance to collide with anything during the initial very short times. 
There exists the possibility of the direct transition from the ballistic regime to the single-file behavior 
\cite{Karger08,KargerHahn96,HahnKarger96}.
Such a tendency becomes more prominent with increasing particle concentration, 
and this has been demonstrated by molecular dynamics simulations \cite{KargerHahn96,HahnKarger96}.
In order to describe such a situation, 
one consider the generalized Langevin equation 
(\ref{eq:withoutExForce_001})
with $f(t) = 0$ and $\gamma(t) = \lambda_2 t^{-1/2}$, which gives the MSD as
\begin{align}
  R^2 & = 2k_BT t^2 E_{3/2,3}\left(-\lambda_2 t^{3/2}\right),
\label{eq:clossExp_004}
\end{align}
The detailed derivation of 
(\ref{eq:clossExp_004})
 is given in \ref{sec:appendixC}. 
It can be shown easily that~(\ref{eq:clossExp_004}) gives the correct long and short time limits for a SF system 
that goes directly from ballistic motion to the SF regime. 
Simulation of (\ref{eq:clossExp_004}) is given in Figure \ref{fig:SFD02}. 
We also show that the presence of the three regimes, 
namely the ballistic motion, normal diffusion and SFD subdiffusion in Figure \ref{fig:SFD03} based on (\ref{eq:clossExp_004}). 

\section{Concluding Remarks}
\label{sec:conclude}
\noindent
We have studied a class of generalized Langevin equation with and without external force. 
It is found that with a specific choice of external force, namely one which varies as $t^{-\kappa}$, $0< \kappa \leq 1$, 
then the solution of the generalized Langevin equation 
(\ref{eq:withoutExForce_001})
gives a Gaussian process with MSD that has the same asymptotic properties as that of SFD.

We remark that an equivalent way to characterized SFD is to use the probability distribution function $W(x,t)$
which can be obtained as solution to the Fokker-Planck equation. 
Derivation of Fokker-Planck equaion corresponds to the fractional generalized Langevin equation 
(\ref{eq:withoutExForce_001}) is a highly non-trivial problem. Furthermore, it may not be fruitful to consider such an equation since we only use 
(\ref{eq:withoutExForce_001}) to obtain the asymptotic properties of SFD. 
However, it may be interesting to note that
$W(x,t)$ at the short and long time limits are respectively given by solutions to the following Fokker-Planck equations: 
\begin{align}
  \frac{\partial W(x,t)}{\partial t} & = D_\circ \frac{\partial^2 W(x,t)}{\partial x^2},
    \quad t \rightarrow 0,
\label{eq::conclude_001}
\end{align}
and
\begin{align}
  \frac{\partial W(x,t)}{\partial t} & = \frac{F}{2\sqrt{t}} \frac{\partial^2 W(x,t)}{\partial x^2},
    \quad t \rightarrow \infty.
\label{eq::conclude_002}
\end{align}
(\ref{eq::conclude_001})
is just the usual diffusion equation and its solution gives the probability distribution of normal diffusion, 
while 
(\ref{eq::conclude_002})
is similar to the effective Fokker-Planck equation for fractional Brownian motion with Hurst index 
$H=1/4$
\cite{WangLung90}. 
One may heuristically combine these two equations into one:
\begin{subequations}
\label{eq::conclude_003}
\begin{align}
  \frac{\partial W(x,t)}{\partial t} & = \left(\frac{F}{D_\circ\sqrt{t}}\right)^{\rho(t)}D_\circ \frac{\partial^2 W(x,t)}{\partial x^2},
\label{eq::conclude_003a}
\end{align}
where
\begin{align}
  \rho(t)
  \begin{cases}
    0 & \text{if} \ t \rightarrow 0   \\
    1 & \text{if} \ t \rightarrow \infty   
  \end{cases}.
\label{eq::conclude_003b}
\end{align}  
\end{subequations}
(\ref{eq::conclude_003}) has the disadvantage that the intermediate values of $\rho(t)$ is not known. 
One concrete version of Fokker-Planck equation 
that gives the short and long time limiting equations (\ref{eq::conclude_001}) and (\ref{eq::conclude_002})
is the following:
\begin{align}
   \frac{\partial W(x,t)}{\partial t} & = D_\circ \! \sqrt{\pi}
                                          E_{1/2,1}\left(-\frac{2D_\circ}{F}t^{1/2}\right)
                                          \frac{\partial^2 W(x,t)}{\partial x^2},
\label{eq::conclude_004}
\end{align}
which has the required properties since
\begin{align}
   E_{1/2,1}\left(-\frac{2D_\circ}{F}t^{1/2}\right) & \underset{t \to 0}{\sim} \frac{1}{\Gamma(1/2)}
                                                     = \frac{1}{\sqrt{\pi}}
\label{eq::conclude_005}
\end{align}
and
\begin{align}
   E_{1/2,1}\left(-\frac{2D_\circ}{F}t^{1/2}\right) & \underset{t \to \infty}{\sim} 
                                                      \frac{1}{\Gamma(1/2)\frac{2D_\circ}{F}t^{1/2}}
                                                     = \frac{F}{2\!\sqrt{\pi}D_\circ t^{1/2}}.
\label{eq::conclude_006}
\end{align}

Note that (\ref{eq::conclude_004}) is one possible effective Fokker-Planck equation which gives the correct asymptotic
probability distributions for SFD.
Our remarks on the limitation of using MSD to characterize SFD apply to probability distribution as well. 
$W(x,t)$ does not describe SFD uniquely even if we know its values for all intermediate times, 
or the correct Fokker-Planck equation for the process. 
On the other hand, if one has the correct fractional Langevin equation for SFD at all times, 
then it describes the process uniquely.

We also obtain a class of new closed expressions for MSD of SFD by considering the solution of the overdamped case, 
thus providing an alternative expression to that of Brandani \cite{Brandani96}. 
From another special case of generalized fractional Langevin equation (\ref{eq:withoutExForce_001}), 
we derive a closed expression describing MSD of SFD with three regimes, namely ballistic motion, 
normal diffusion and sub-diffusion. 
These closed analytic expressions of MSD are given in terms of Mittag-Leffler functions. 
Finally, we remark that despite of numerous studies carried out on SFD, 
there are very few results which deal with possible concrete realizations of SFD as a specific stochastic process. 
In a recent work, we proposed one such possible realization of SFD as the step fractional Brownian motion, 
which has the flexibility of giving the correct description to SFD with two and three regimes 
\cite{LimTeo09}. 

\appendix

\section{}
\label{sec:appendixA}
\noindent
Here we give some properties of $G(t)$ which are required for obtaining the asymptotic values of MSD. 
We have the following properties of fractional integrals and derivatives of $G(t)$:
\begin{subequations}
\label{eq:appendixA_001}
\begin{align}
  I^{1-\mu} G(t) & = I^{1-\mu}G_\circ(t) \nonumber \\
                 & \quad  
                 + \sum_{n=1}^\infty \left(-\lambda_1\right)^n\int_0^t du I^{1-\mu} G_\circ(t - u) G_\circ^{*n}(u)
\label{eq:appendixA_001a}
\end{align}
and
\begin{align}
  D^\phi G(t) & = D^\phi G_\circ(t) \nonumber \\
              & \quad
                   + \sum_{n=1}^\infty \left(-\lambda_1\right)^n\int_0^t du D^\phi G_\circ(t - u) G_\circ^{*n}(u).
\label{eq:appendixA_001b}
\end{align}  
\end{subequations}
From (\ref{eq:withExForce_005a}) the following can be calculated
\begin{subequations}
\label{eq:appendixA_002}
\begin{align}
  D^\alpha G_\circ(t) &= E_{\alpha - \gamma +1,1}\left(-\lambda_2 t^{\alpha-\gamma+1}\right)  
\label{eq:appendixA_002a}
\\
  I^{1-\alpha} G_\circ(t) &= tE_{\alpha - \gamma +1,2}\left(-\lambda_2 t^{\alpha-\gamma+1}\right)  
\label{eq:appendixA_002b}
\\
  I^{1-\kappa} G_\circ(t) &= t^{\alpha-\kappa+1}E_{\alpha - \gamma +1,\alpha-\kappa+2}\left(-\lambda_2 t^{\alpha-\gamma+1}\right).  
\label{eq:appendixA_002c}
\end{align}
\end{subequations}
In addition, one needs to consider the following:
\begin{align}
  \int_0^t du G_\circ(u) &= \int_0^t du u^{\alpha}
                       E_{\alpha - \gamma +1,\alpha+1}\left(-\lambda_2 u^{\alpha-\gamma+1}\right)  \nonumber\\
                    & = t^{\alpha+1}E_{\alpha - \gamma +1,\alpha+2}\left(-\lambda_2 t^{\alpha-\gamma+1}\right). 
\label{eq:appendixA_003}
\end{align}

\section{}
\label{sec:appendixB}
\noindent
We give the derivation of the expression for variance (\ref{eq:withExForce_015}). 
From (\ref{eq:withoutExForce_008}), the Laplace transform of the convolution
\begin{align}
  A(t) & = \int_0^t du C(t - u)G(u)
\label{eq:appendixB_001}
\end{align}
is given by
\begin{align}
  \tilde{A}(s) &= \tilde{C}(s)\tilde{G}(s).
\label{eq:appendixB_002}
\end{align}
From the fluctuation dissipation relation we have  
$C(t) = k_BT\gamma(t)$, 
and express according to definition (\ref{eq:withoutExForce_007}), we have
\begin{align}
  \tilde{A} &= k_BT\frac{\tilde{\gamma}(s)}{s^{\alpha+1} + \tilde{\gamma}(s) s}
             = \frac{k_BT}{s}\left[1 - \frac{s^\alpha}{s^\alpha + \tilde{\gamma}(s)}\right].
\label{eq:appendixB_003}
\end{align}
Replace the convolution term in 
(\ref{eq:withoutExForce_008})
 by the inverse Laplace transform of (\ref{eq:appendixB_003}) we get (\ref{eq:withExForce_015}).

\section{}
\label{sec:appendixC}
\noindent
Consider the fractional generalized Langevin equation 
(\ref{eq:withoutExForce_001}) and (\ref{eq:withoutExForce_002})  
with $f(t) = 0$. 
The solution for the position process is given by
\begin{align}
  x(t) &= x_\circ + \frac{v_\circ}{\Gamma(1 - \alpha}\int_0^t (t - u)^{-\alpha} G(u) du \nonumber \\
       & \quad + \int_0^t G(t - u)\xi(u) du,
\label{eq:appendixC_001}
\end{align}
where $G(t)$ is the inverse Laplace transform of 
\begin{align}
  \tilde{G}(s) &= \frac{1}{s^{\alpha+1} + \tilde{\gamma}(s) s}.
\label{eq:appendixC_002}
\end{align}
For $\gamma(t) = \lambda_2 \frac{t^{-\gamma}}{\Gamma(1 - \gamma)}$, one gets
\begin{align}
  G(t) &= t^\alpha E_{\alpha - \gamma +1,\alpha +1}\left(-\lambda_2 t^{\alpha - \gamma +1}\right).
\label{eq:appendixC_003}
\end{align}
The mean and variance of $x(t)$ are respectively
\begin{align}
  \bar{x} &= x_\circ + v_\circ I^{1-\alpha}G(t)
\label{eq:appendixC_004}
\end{align}
and
\begin{align}
  \left<\left(x(t) - \bar{x}\right)^2\right> &= 2k_BT\left[
                                                       \int_0^t du G(t)
                                                       - \int_0^t du G(t)D^\alpha G(u)
                                                     \right].
\label{eq:appendixC_005}
\end{align}
Now using the relation $R^2 = \left(\bar{x} - x_\circ\right)^2 + \sigma^2$, 
one sees that the MSD $R^2$ can have a close form provided the second term of the variance cancels out the term 
$\left(\bar{x} - x_\circ\right)^2$. 
This is possible if $\alpha = 1$.  
We thus obtain
\begin{align}
  R^2 &= 2k_BT\int_0^t du G(u)
       = 2k_BT\int_0^t du uE_{2-\gamma,2}\left(-\lambda_2 u^{2-\gamma}\right) \nonumber \\
      & = 2k_BT t^2 E_{2-\gamma,3}\left(-\lambda_2 t^{2-\gamma}\right).
\label{eq:appendixC_006}
\end{align}
When $\gamma = 1/2$, one gets 
\begin{align}
  R^2 = 2k_BT t^2 E_{3/2,3}\left(-\lambda_2 t^{3/2}\right).
\label{eq:appendixC_007}
\end{align}

\bibliographystyle{plain}
\bibliography{biblioSFD}

\end{document}